\documentclass[sigconf]{acmart}

\usepackage{caption}
\usepackage{subcaption}

\usepackage[ruled,linesnumbered]{algorithm2e}
\let\oldnl\nl
\newcommand{\nonl}{\renewcommand{\nl}{\let\nl\oldnl}}
\SetKwInput{KwData}{Input}
\SetKwInput{KwResult}{Output}
\DontPrintSemicolon

\AtBeginDocument{%
  \providecommand\BibTeX{{%
    \normalfont B\kern-0.5em{\scshape i\kern-0.25em b}\kern-0.8em\TeX}}}

\setcopyright{acmcopyright}
\copyrightyear{2022}
\acmYear{2022}
\acmDOI{XXXXXXX.XXXXXXX}

\acmConference[Conference acronym 'XX]{Make sure to enter the correct
  conference title from your rights confirmation email}{June 03--05,
  2022}{Woodstock, NY}
\acmPrice{15.00}
\acmISBN{978-1-4503-XXXX-X/18/06}

\begin{document}

\title{Rache: Radix-additive caching for homomorphic encryption}

\author{Dongfang Zhao}
\email{dzhao@unr.edu}
\affiliation{%
  \institution{University of Nevada}
  \city{Reno}
  \country{United States}
}

\begin{abstract}
One of the biggest concerns for many applications in cloud computing lies in data privacy.
A potential solution to this problem is homomorphic encryption (HE),
which supports certain operations directly over the ciphertexts.
Conventional HE schemes, however, 
exhibit significant performance overhead and are hardly applicable to real-world applications.
This paper presents Rache, a caching optimization for accelerating the performance of HE schemes.
The key insights of Rache include (i) caching some homomorphic ciphertexts before encrypting the large volume of plaintexts;
(ii) expanding the plaintexts into a summation of powers of radixes; and 
(iii) constructing the ciphertexts with only homomorphic addition.
The extensive evaluation shows that Rache exhibits almost linear scalability and outperforms Paillier by orders of magnitude.
\end{abstract}

\settopmatter{printfolios=true}
\maketitle

\section{Introduction}

While increasingly more applications are deployed on the public cloud,
one of the biggest concerns lies in data privacy,
especially for those applications that usually touch on sensitive data in the fields of public health~\cite{tkanwal_cluster21}, bioinformatics~\cite{xzhu_tdsc21}, financial services~\cite{wkuan_clsr18}, among others.
While various encryption schemes (e.g., AES~\cite{aes}, RSA~\cite{rsa}) exist,
it would defeat the purpose and advantage of cloud computing if the users (i) encrypt the sensitive data, (ii) upload the data to the cloud, (iii) download a portion of data of interests to a local disk, and (iv) decrypt the ciphertexts for some computation,
because the cloud in this case works merely as a remote backup storage without any \textit{computing} functionality.

One solution to the data privacy problems in cloud computing is to design specific encryption schemes such that the ciphertexts stored on the cloud can perform certain computations.
Technically, such an encryption scheme is called a \textit{homomorphic encryption} (HE) scheme.
Although many HE schemes support only primitive arithmetic operations such as addition and multiplication,
it turns out that many commonly-used operations can be implemented using addition and/or multiplication~\cite{cgentry_stoc09}.
The first type of HE schemes, e.g., Symmetria~\cite{symmetria_vldb20}, are implemented as a symmetric operation for the scenarios where a secret key can be securely shared among parties,
which is not always possible in cloud computing.
The second type of HE schemes, e.g., Paillier~\cite{ppail_eurocrypt99},
are implemented as an asymmetric operation that overcomes the limitation of a symmetric one,
and yet introduces significant performance overhead,
making it impractical to encode a large volume of sensitive data.
Although a hybrid scheme can be used to encrypt the secret key using asymmetric encryption one time,
this process works only for a single session and the asymmetric encryption would have to be invoked many times in a production environment.

This paper presents a new caching method,
namely radix-additive caching for homomorphic encryption (Rache), 
for accelerating the performance of asymmetric homomorphic encryption represented by Paillier~\cite{ppail_eurocrypt99}.
The key insights of Rache include:
(i) precomputing and caching some homomorphic ciphertexts before encrypting the large volume of plaintexts;
(ii) expanding a plaintext into a summation of additive radix entries; and
(iii) constructing the ciphertexts with only homomorphic addition (without touching on any homomorphic encryption).
The third insight is inspired by our conjecture that a homomorphic addition is much cheaper than a homomorphic encryption,
which we will justify in~\S\ref{sec:eval_micro}.

This paper makes the following technical contributions.
\begin{itemize}
    \item We present Rache, a new caching method for accelerating homomorphic encryption. 
    (\S\S\ref{sec:rache_desc},~\ref{sec:rache_correct})
    \item We analyze the theoretical complexity of Rache and derive the worst-case optimal radix. 
    (\S\S\ref{sec:rache_complexity},~\ref{sec:rache_radix})
    \item We implement a system prototype of Rache with C and OpenSSL.
    (\S\ref{sec:eval_impl})
\end{itemize}

We evaluate Rache with three benchmarks (e.g. TPC-H~\cite{tpch3}) and three real-world applications (e.g., Covid-19~\cite{covid19data}) on CloudLab~\cite{cloudlab}.
Experimental results show that:
\begin{itemize}
    \item Rache outperforms Paillier by orders of magnitude on all of the six workloads;
    (\S\S\ref{sec:eval_benchmark},~\ref{sec:eval_app})
    \item Rache exhibits almost linear scalability in both weak- and strong-scaling experiments on up to 32 cores;
    (\S\ref{sec:eval_scale})

    \item Rache consumes less memory than Paillier by up to 12\%.
    (\S\ref{sec:eval_memory})
\end{itemize}

The source code of Rache will be released under Apache License 2.0.

\section{Background and Related Work}

\textit{Homomorphic encryption} (HE) is a specific type of encryption where certain operations between operands can be performed directly on the ciphertexts.
For example, if an HE scheme $he(\cdot)$ is additive,
then the plaintexts with $+$ operations can be translated into a homomorphic addition $\oplus$ on the ciphertexts.
Formally, if $a$ and $b$ are plaintexts, then the following holds:
\[
he(a) \oplus he(b) = he(a + b).
\]
As a concrete example, let $he(x) = 2^x$, and we temporarily release the security requirement of $he(\cdot)$.
In this case, $he(a+b) = 2^{a+b} = 2^a \times 2^b = he(a) \times he(b)$,
meaning that $\oplus$ is the arithmetic multiplication $\times$.

An HE scheme that supports addition is said to be \textit{additive}.
Popular additive HE schemes include Symmetria~\cite{symmetria_vldb20} and Paillier~\cite{ppail_eurocrypt99}.
The former is applied to symmetric encryption,
meaning that a single secret key is used to both encrypt and decrypt the messages.
The latter is applied to asymmetric encryption,
where a pair of public and private keys are used for encryption and decryption.
Due to the expensive arithmetical operations performed by the asymmetric encryption,
Paillier is orders of magnitude slower than Symmetria.
However, Paillier is particularly useful when there is no secure channel to share the secret key among users,
which is required by symmetric encryption schemes.
One notable extension of Symmetria is to incrementally encrypt plaintexts with the additive homomorphism property, as illustrated in~\cite{dzhao_inche}.

An HE scheme that supports multiplication is said to be \textit{multiplicative}.
Symmetria~\cite{symmetria_vldb20} is also multiplicative using a distinct scheme than the one for addition.
Other well-known multiplicative HE schemes include RSA~\cite{rsa} and ElGamal~\cite{elgamal_tit85}.
Similarly, a multiplicative HE scheme guarantees the following equality,
\[
he(a) \otimes he(b) = he(a \times b),
\]
where $\otimes$ denotes the homomorphic multiplication over the ciphertexts.

An HE scheme that supports both addition and multiplication is called a \textit{fully HE scheme}.
This requirement should not be confused with specific addition and multiplication parameters, such as Symmetria~\cite{symmetria_vldb20} and NTRU~\cite{ntru}.
That is, the addition and multiplication must be supported homomorphically under exactly the same scheme $he(\cdot)$:
\[\displaystyle
\begin{cases}
    he(a) \oplus he(b) = he(a + b) \\
    he(a) \otimes he(b) = he(a \times b)
\end{cases}
\]
It turned out to be extremely hard to construct fully HE schemes until Gentry~\cite{cgentry_stoc09} demonstrated such a scheme using lattice theory.
The main issue with fully HE schemes is their performance;
although extensive research and development have been carried out, 
current implementations incur impractical overhead for most real-world applications.  
One notable attempt to boost the performance of fully HE schemes is to distribute the computation~\cite{jyin_bigdata15}.
Two popular open-source libraries of fully HE schemes are IBM HElib~\cite{helib} and Microsoft SEAL~\cite{sealcrypto}.
The Rache scheme presented in this paper needs only the additive property and does not require a fully HE.

In a positional numeral system, a number $x$ can be written as a summation of terms,
each of which is a product of two factors---one is the integral power of \textit{radix} $r$ and the other is the coefficient ranging from 0 to $r-1$.
Formally,
\begin{equation}\label{eq:radix}\displaystyle
x = idx_0 \cdot r^0 + idx_1 \cdot r^1 + \dots + idx_k \cdot r^k,    
\end{equation}
where $idx_i$ ($0 \le i < r$) indicates the coefficient of a specific \textit{radix entry}.
Eq.~\ref{eq:radix} can be further expanded into an expression with only additions:
\begin{equation}\label{eq:additive}
\displaystyle
\begin{split}
    x 
    &= \underbrace{\left(r^0 + r^0 + \dots \right)}_{idx_0} + \underbrace{\left(r^1 + r^1 + \dots\right)}_{idx_1} + \dots + \underbrace{\left(r^k + r^k + \dots\right)}_{idx_k} \\
    &= \sum_{i=0}^k \sum_{j=1}^{idx_i} r^i,
\end{split}
\end{equation}
where $idx_i < r$. 
This purely additive form in Eq.~\ref{eq:additive} allows us to apply additive homomorphic encryption \textit{inside} the radix entries rather than the original number $x$,
as we will start discussing in the next section.

\section{Radix Caching for Homomorphic Encryption}

\subsection{Description}
\label{sec:rache_desc}

Alg.~\ref{alg:rache} formalizes the encoding procedure with C-like pseudocode.
Lines 1--4 initialize the cached entries of the integral powers of radix $r$ for future construction of ciphertexts.
Lines 5--10 encode the plaintexts,
each of which is computed directly over the cached ciphertexts that are initialized at the beginning of the algorithm.
We will discuss the algorithm's correctness, complexity, and choice of $r$ in the remainder of this section.

\begin{algorithm}
\caption{Radix caching in homomorphic encryption}\label{alg:rache}
\KwData{
An array of plaintexts $Ptxt[]$ of length $n$; 
A homomorphic encryption scheme $he(\cdot)$ s.t. $\forall a_i \in Ptxt[], \bigoplus_i he(a_i) = he(\sum_i a_i)$;
Radix $r$;
}
\KwResult{
An array of ciphertexts $Ctxt[]$ s.t. $\forall i \in \mathbb{Z}$, $0 \le i < n$, $Ctxt[i] = he\left(Ptxt[i]\right)$;
}
 
\nonl \;
\nonl // Initialization\;
$m \gets$ maximal value of $Ptxt[]$\;
\For{$i = 0; i <= \lfloor \log_r m \rfloor; i++$}{
    $radixes[i] \gets he(r^i)$\;
}
 
\nonl \;
\nonl // Encoding \;
\For{i = 0; i < n; i++}{
    \For{$j = 0; j <= \lfloor \log_r m \rfloor; j++$}{
        $idx[j] \gets$ $(Ptxt[i]$ / $r^j)$ \% $r$\;
    }
    \nonl // $Ptxt[i] = \sum_j idx[j]\times r^j$ \;
    $Ctxt[i] = \bigoplus_{k=0}^{\lfloor \log_r m \rfloor} \bigoplus_{j=1}^{idx[k]} radixes[k]$\;
}
\end{algorithm}

\subsection{Correctness}
\label{sec:rache_correct}

We denote $\bigoplus$ the homomorphic summation over the ciphertexts.
The correctness of Alg.~\ref{alg:rache} can be verified by direct computation as follows.
\[\displaystyle
\begin{split}
Ctxt[i]
&= \bigoplus_{k=0}^{\lfloor \log_r m \rfloor} \bigoplus_{j=1}^{idx[k]} radixes[k] \\
&= \bigoplus_{k=0}^{\lfloor \log_r m \rfloor} \bigoplus_{j=1}^{idx[k]} he(r^k) \\
&= \bigoplus_{k=0}^{\lfloor \log_r m \rfloor} he\left( \sum_{j=1}^{idx[k]} r^k \right) \\
&= \bigoplus_{k=0}^{\lfloor \log_r m \rfloor} he\left( idx[k] \times r^k \right) \\
&= \bigoplus_{k=0}^{\lfloor \log_r m \rfloor} he\left( (Ptxt[i] / r^k) \% r \times r^k \right) \\
&= he\left( \sum_{k=0}^{\lfloor \log_r m \rfloor} (Ptxt[i] / r^k) \% r \times r^k \right) \\
&= he\left( Ptxt[i] \right)
\end{split}
\]

The first equality is due to Line 9 of Alg.~\ref{alg:rache}.
The second equality is due to Line 3 of Alg.~\ref{alg:rache}.
The third equality is due to the definition of homomorphic encryption.
The fourth equality is due to the fact that variable $j$ does not show up in the term $r^k$.
The fifth equality is due to Line 7 of Alg.~\ref{alg:rache}.
The sixth equality is, again, due to the definition of homomorphic encryption.
The last equality is due to the definition of radix expansion.

\subsection{Complexity}
\label{sec:rache_complexity}

We denote $w$ the time cost of homomorphically encrypting a number.
We denote $h$ the time cost of homomorphically adding two ciphertexts.
We will soon see that in practice $w$ is much larger (i.e., more than two orders of magnitude) than $h$ in~\S\ref{sec:eval_micro}.
Line 1 takes $\mathcal{O}(1)$ if we assume the system caches the maximal plaintext when reading $Ptxt[]$ into the memory.
Lines 2--4 take $\mathcal{O}(w\log m)$.
Lines 6--8 take $\mathcal{O}(\log m)$.
Lines 9 takes $\mathcal{O}(rh\log m)$.
Therefore, Lines 5--10 take $\mathcal{O}(n\cdot (\log m + rh\log m)) = \mathcal{O}(rhn\log m)$.
The overall time cost of Alg.~\ref{alg:rache} is thus $\mathcal{O}(w\log m + rhn\log m)$.

In practice, $r$ is a small number (in fact, the following section will show that $r = 2$ is an optimal radix in the worst case).
The cost of homomorphic addition $h$ is on par with that of regular arithmetic operation and can be considered as 1.
The factor $\log m$ is a small number as well; for example, for encrypting a 1,000,000,000 number, 
30 is good enough with radix 2.
Consequently, a more practical upper bound of Alg.~\ref{alg:rache} is $\mathcal{O}(cw + 2cn)$, 
where $c$ denotes a constant.
That is, Alg.~\ref{alg:rache} costs time that is about constant folds of the homomorphic encryption and the overall number of ciphertexts.

Recall that the time cost of Paillier is simply $\mathcal{O}(nw)$,
which is the multiplication of the homomorphic encryption cost and the overall number of ciphertexts.
We thus expect Alg.~\ref{alg:rache} will outperform Paillier by orders of magnitude.
Before we demonstrate the performance superiority of Alg.~\ref{alg:rache}, i.e., Rache,
we conclude this section with a discussion on the optimal choice of radix in the worst case.

\subsection{Optimal Radix}
\label{sec:rache_radix}

Let $m \ge 2$ denote the maximal number to be encrypted in the application.
Let $r \ge 2$ denote the radix or base of the homomorphic encryption.
Obviously, given an arbitrary number $x$, where $0 \le x \le m$,
there are $k + 1$ radix entries:
$r^0$, $r^1$, $\dots$, $r^k$,
where $k = \lfloor \log_r^m \rfloor$.
Let $0 \le \kappa \le k$.
In the worst case, each $r^{\kappa}$ radix-entry incurs $r - 2$ times of homomorphic addition,
i.e., when computing $(r - 1) \cdot x^{\kappa}$.
Since one more homomorphic addition needs to be taken for the summation of each radix,
the overall times of homomorphic addition, in the worst case when $m$ is one less than the next integral power of $r$ (i.e., $\lfloor \log_r^m \rfloor = \log_r^{m+1} - 1$), is 
\[\displaystyle
\begin{split}
f(r)&= (r-2)(k+1) + k \\
    &= (r-1)k + r - 2 \\
    &= (r-1)(\log_r^{m+1} - 1) + r - 2 \\
    &= (r-1)\log_r^{m+1} - 1.
\end{split}
\]
We will find out the optimal $r$ that minimizes $f(r)$.

We take the first-order derivative of $f(r)$ as follows.
\[\displaystyle
\begin{split}
f'(r) 
&= \frac{d}{dr} f(r) \\
&= \frac{d}{dr} \left( (r - 1)\log_r^{m+1} - 1 \right) \\
&= \frac{d(r-1)}{dr}\cdot \log_r^{m+1} + (r-1)\cdot \frac{d}{dr} \left( \log_r^{m+1} \right) \\
&= 1\cdot \log_r^{m+1} + (r-1)\cdot \frac{d}{dr} \left(\frac{\ln (m+1)}{\ln r} \right) \\
&= \log_r^{m+1} + (r-1)\ln (m+1) \cdot \frac{d}{dr} \left( \left(\ln r \right)^{-1} \right) \\
&= \log_x^{m+1} + (r-1)\ln (m+1) \cdot (-1) \cdot (\ln r)^{-2} \cdot \frac{1}{r} \\
&= \ln (m+1) \cdot (\ln r)^{-1} - \ln (m+1) \cdot (\ln r)^{-2} \cdot \frac{r-1}{r} \\
&= \ln (m+1) \cdot (\ln r)^{-2} \cdot r^{-1} \cdot \left( r\ln r - r + 1 \right).
\end{split}
\]
The stationary point is therefore the solution to $f'(r) = 0$,
\[\displaystyle
g(r) = r\ln r - r + 1 = 0,
\]
which yields $r = 1$.
Since we require $r \ge 2$, we need to find another qualified radix.
First, we calculate $g(2)$:
\[\displaystyle
g(2) = 2\ln 2 - 2 + 1 \ge 2\times 0.69 - 1 > 0.
\]
Then, let $r \ge 3$, therefore $\ln r > 1$, which yields:
\[\displaystyle
\left.g(r)\right\rvert_{r \ge 3} = r\ln r - r + 1 = r(\ln r - 1) + 1 > 0.
\]
Note that by definition, the following equation holds:
\[\displaystyle
f'(r) = \ln (m+1) \cdot (\ln r)^{-2} \cdot r^{-1} \cdot g(r).
\]
If we assume $m \ge 2$, then $\ln (m+1) > 0$.
Both $(\ln r)^{-2}$ and $r^{-1}$ factors are obviously positive.
Therefore, $f'(r)$ is always positive, meaning that $f(r)$ is a monotonically increasing function.
It follows that the minimal qualified radix $r = 2$ leads to the minimum number of homomorphic additions.

\section{Evaluation}
\label{sec:eval}

\subsection{Implementation}
\label{sec:eval_impl}

We implement Rache with C and three key libraries: Open-MPI~\cite{openmpi} (C binding), OpenSSL~\cite{openssl_github}, and homomorphic-c~\cite{homoc_github}.
Specifically, the arbitrarily large numbers are managed with the BIGNUM structure.
The baseline homomorphic encryption scheme is Paillier~\cite{ppail_eurocrypt99},
which is also implemented with C and OpenSSL.
It is easy for memory to leak in C,
our implementation makes sure that all memory allocation is appropriately released (see~\S\ref{sec:eval_memory} for quantitative evaluation).
At the time of writing this paper, the implementation consists of 11,839 lines of code.

\subsection{Experimental Setup}
All experiments were carried out on the CloudLab testbed~\cite{cloudlab}.
We use the \texttt{c6420} instance,
which is equipped with Intel Xeon Gold 6142 CPUs at 2.6 GHz, 384 GB ECC DDR4-2666 memory, and 	
two Seagate 1~TB 7200 RPM 6G SATA HDDs.
Each node has 32 physical cores and supports 64 hyperthreads. 
The operating system image is Ubuntu 20.04.3 LTS.
The system is installed with the following notable libraries:
gcc 9.3.0, Open-MPI 4.0.3, and OpenSSL 1.1.1.

Our baseline encryption scheme is Paillier~\cite{ppail_eurocrypt99},
which is implemented in C and the OpenSSL library~\cite{openssl_github}.
Most experiments adopt a strong-scaling mechanism, 
meaning that the given workload is split by a variety of cores,
ranging from 1 to 32.
We have evaluated Rache with three benchmarks and three real-world applications.

\begin{itemize}

\item The first benchmark is a microbenchmark to quantify the cost of homomorphic encryption and homomorphic addition, respectively.
For the former, a sequence of integers [0, 32,768) are homomorphically encrypted;
for the latter, the ciphertexts stored at radix entries are homomorphically summed up in a round-robin fashion 32,768 times.

\item The second benchmark is TPC-H ver. 3.0.0~\cite{tpch3}, a standard relational database benchmark.
TPC-H allows the user to specify the scales of the generated data;
in this paper we set scale as one, resulting in about one gigabyte of data.
We will focus on the \textit{part} table, which consists of 200,000 tuples.

\item The third benchmark is a dynamic set of random numbers used in INCHE~\cite{dzhao_inche} for homomorphic encryption.
This benchmark is mainly used for the purpose of weak scaling, 
allowing for the scalability test ranging between 1,024 and 32,768 numbers.

\item The first application is the U.S. national Covid-19 statistic from April 2020 to March 2021~\cite{covid19data}.
The data set has 341 days of 16 metrics, such as \textit{death increase}, \textit{positive increase}, and \textit{hospitalized increase}.

\item The second application is the human genome reference 38~\cite{hg_data},
commonly known as \textit{hg38},
which is comprised of 34,424 rows of five singular attributes such as \textit{transcription positions}, \textit{coding regions}, and \textit{number of exons}, last updated in March 2020.

\item The third application is the history of Bitcoin trade volume~\cite{bitcoin_trade} since it was first exchanged in the public in February 2013.
The data consists of the accumulated Bitcoin exchange on a 3-day basis from February 2013 to January 2022,
totaling 1,086 large numbers.

\end{itemize}

We repeat every performance experiment multiple times and report the averages and standard errors.
Radix is set two in all experiments.

\subsection{Benchmarks}
\label{sec:eval_benchmark}

\subsubsection{Micro Benchmark for Encryption and Addition}
\label{sec:eval_micro}

The whole idea of Rache is built upon the assumption that homomorphic addition is a much cheaper operation than homomorphic encryption.
Our first experiment, therefore, tries to confirm this assumption.
The micro-benchmark carries out $n = 32,768$ operations for homomorphic encryption and homomorphic addition, respectively.
Specifically, for encryption, the operation is $he(i)$, $0 \le i < n$;
for addition, the operation is 
\[\displaystyle
he(\lfloor \log i \rfloor) \oplus he((\lfloor \log i \rfloor + 1) \% ( \lfloor \log n \rfloor + 1)),
\] 
where $\oplus$ denotes the homomorphic addition and $\%$ denotes the modular operation.

\begin{figure}[!t]
    \centering
    \includegraphics[width=\columnwidth]{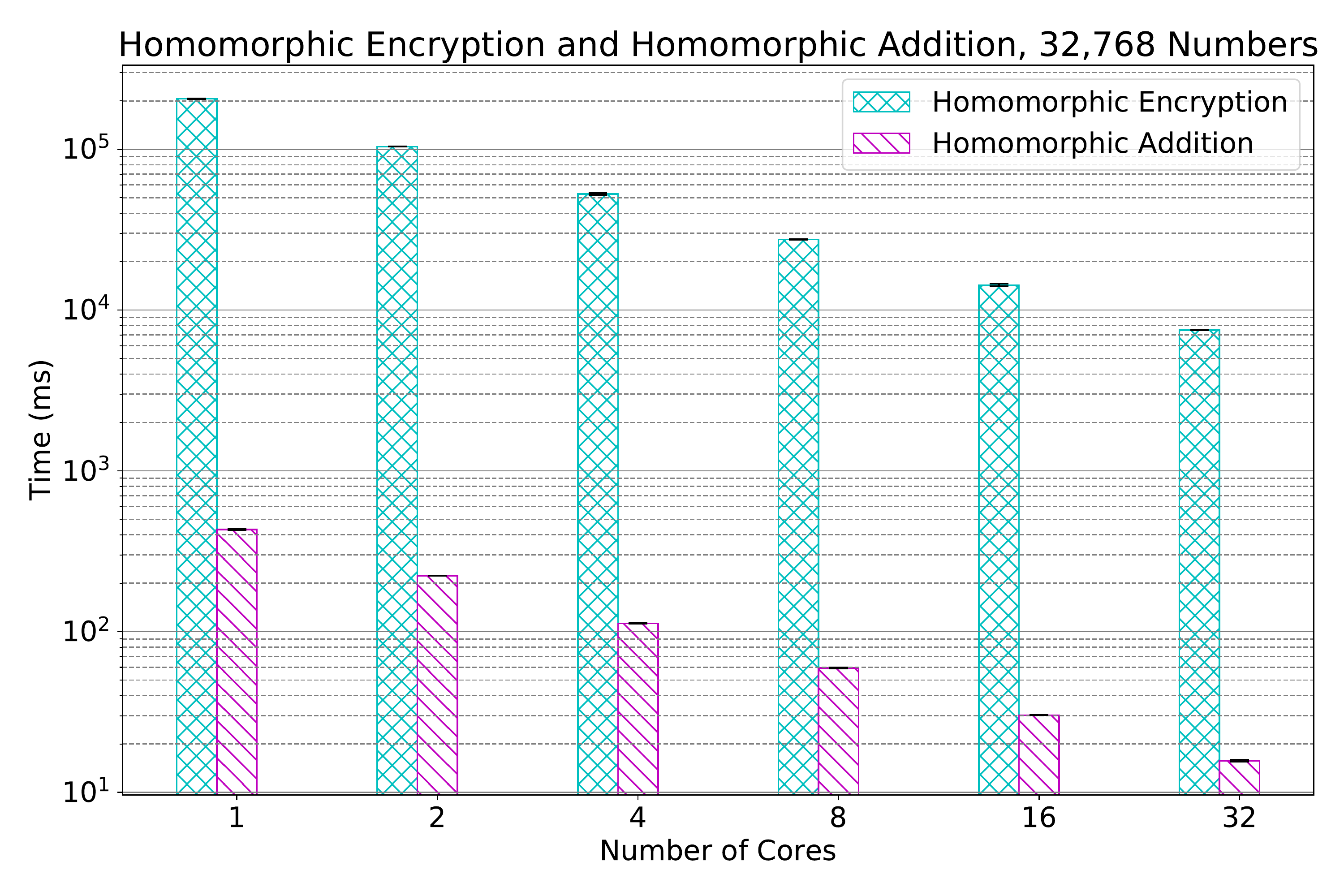}
    \caption{Performance of homomorphic encryption and homomorphic addition.}
    \label{fig:micro}
\end{figure}

Fig.~\ref{fig:micro} shows that the homomorphic addition is a much cheaper operation than homomorphic encryption.
Regardless of the number of available cores,
homomorphic encryption takes more than two orders of magnitude time than homomorphic addition.
Indeed, one Rache encryption typically involves multiple homomorphic additions (plus the constant initialization cost).
The question then becomes whether the multiplication of homomorphic additions incurred by Rache still outperforms the original homomorphic encryption.
The answer is yes, as demonstrated by the following experiments.

\subsubsection{TPC-H}

\begin{figure}[!t]
    \centering
    \includegraphics[width=\columnwidth]{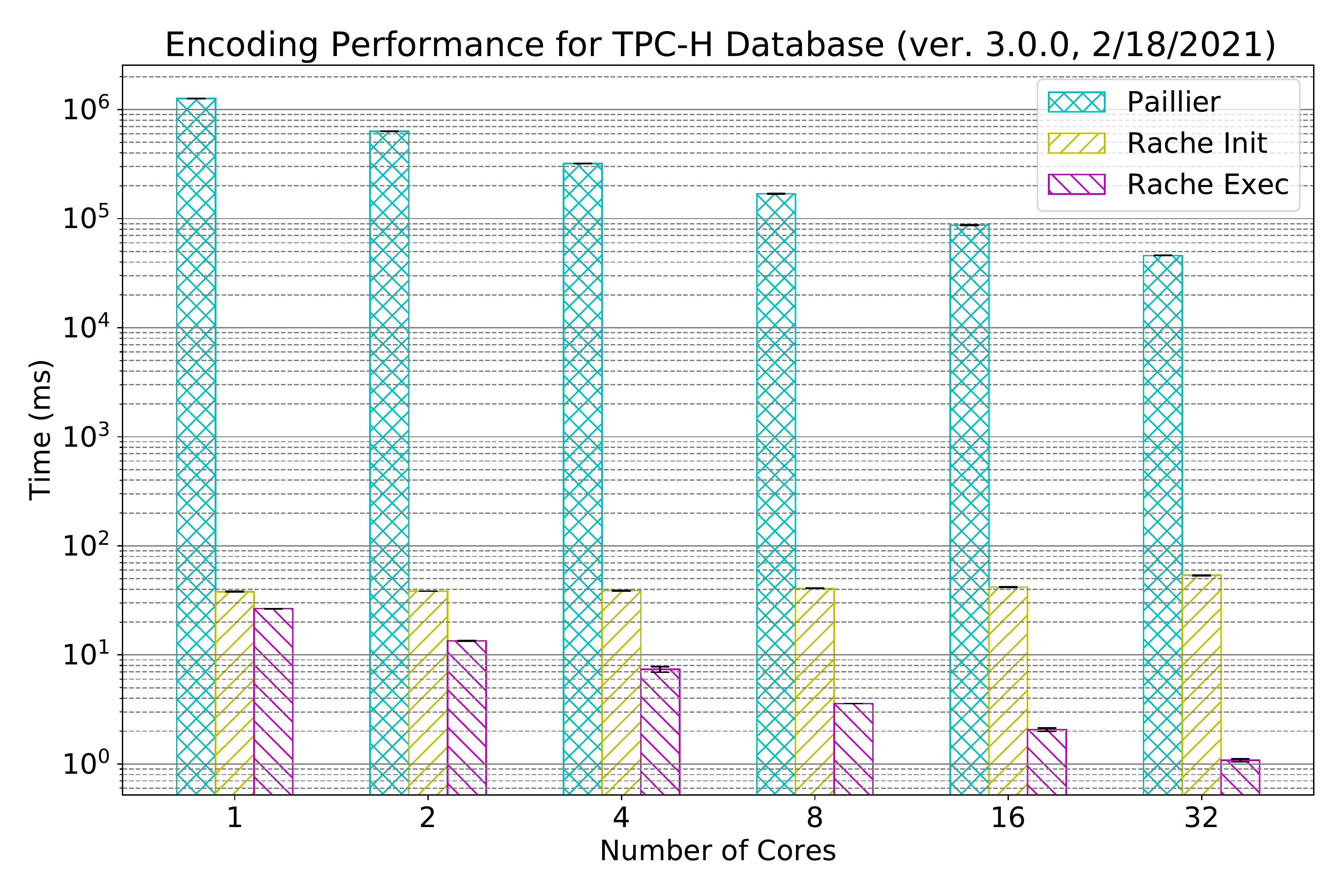}
    \caption{Performance comparison on the TPC-H benchmark.}
    \label{fig:tpch}
\end{figure}

We report Rache's performance of encoding the TPC-H benchmark~\cite{tpch3} in Fig.~\ref{fig:tpch}.
We report the execution time of initializing the radixes and that of encoding with radix cache, respectively.
The former is referred to as \textit{Rache Init} and the latter as \textit{Rache Exec} in the figure (and also in other experiments to be discussed).

Both Rache and Paillier exhibit good (strong) scalability due to the data parallelism from the message passing interface (MPI).
The initialization time of Rache is roughly flattened,
showing a marginal increase when more cores are involved due to the inter-process communication (IPC) overhead.
The overhead is under two orders of magnitude when 32 cores are exploited.

We observe that Rache consistently outperforms Paillier by more than four orders of magnitudes at all scales.
The huge performance gap (even larger than most of the other experiments to be discussed) is partially due to the dataset itself:
The \textit{part} table in TPC-H has relatively small numbers (max 21) such that many of the new numbers to be encoded by Rache can be quickly (homomorphically) constructed by the cached ciphertexts.
We will see how the plaintext affects Rache's performance in the following sections.
We start with a more general setup: encoding a set of random numbers instead of small numbers.

\subsubsection{INCHE Random Numbers}

\begin{figure}[!t]
    \centering
    \includegraphics[width=\columnwidth]{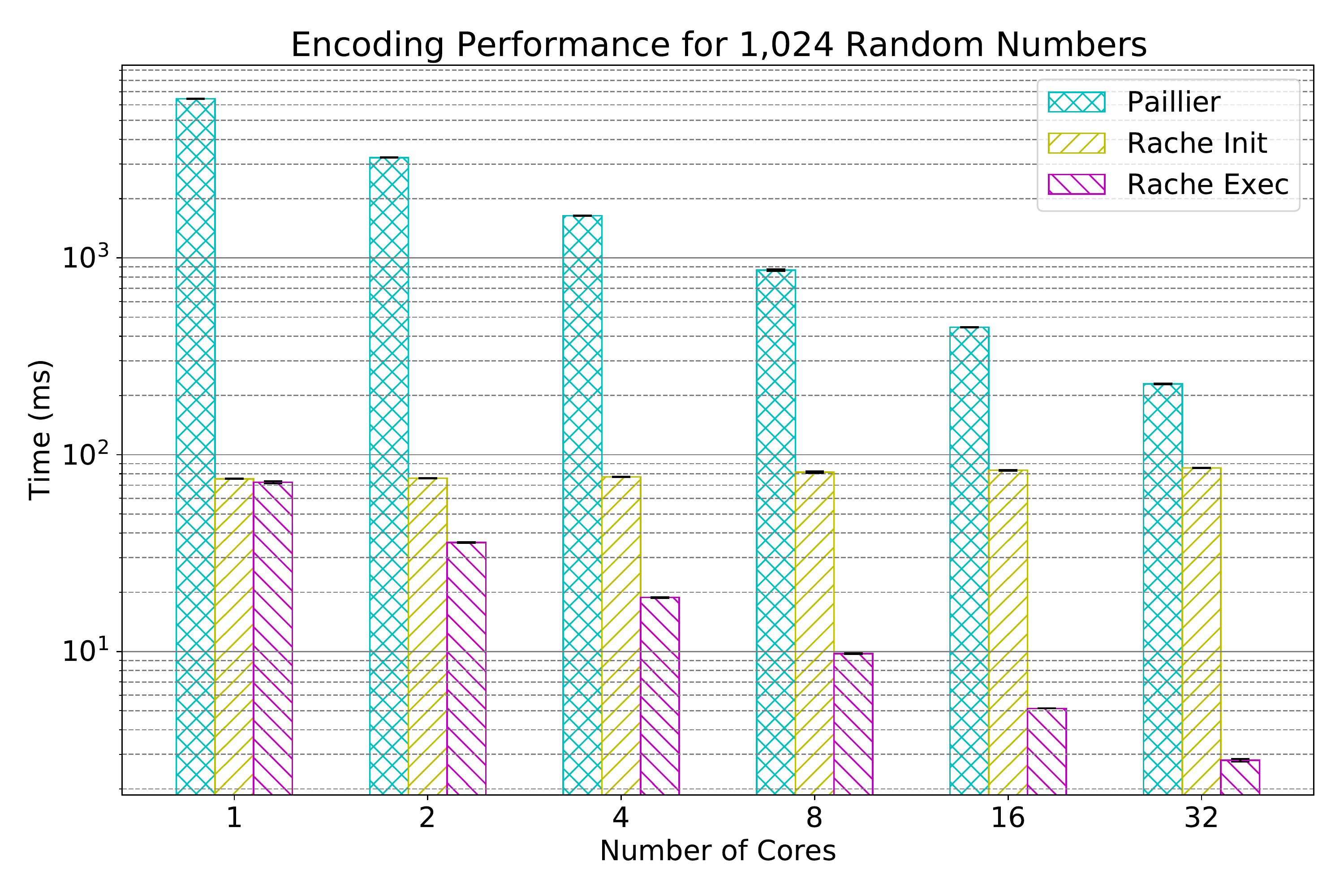}
    \caption{Encoding Performance on Random Numbers.}
    \label{fig:fixed_workload}
\end{figure}

For random numbers, we take the same approach used in~\cite{dzhao_inche}.
Essentially, $n$ random numbers are generated in a uniform distribution by modular $n$.
We report the results in Fig.~\ref{fig:fixed_workload}.
The Rache overhead stays roughly constant for different numbers of cores, but not as low as TPC-H.
This is because the largest number in this INCHE dataset is 1,024,
which requires more radixes to be initialized.
Despite the overhead, we observe that Rache's encoding time is about two orders of magnitude lower than Paillier at all scales.

It should be noted that, however, the \textit{Rache Init} overhead is a one-time thing.
With a larger number of plaintexts,
the overhead does not change as long as the maximal number is unchanged.
We will see this in the following weak-scaling experiment,
i.e., increasing the workloads.

\subsection{Scalability}
\label{sec:eval_scale}

We evaluate the scalability of Rache in this section.
For generality, we focus on the INCHE dataset of random numbers rather than, say, 
specific benchmarks or applications,
which will be the emphases of consequent sections.

\begin{figure}[!t]
    \centering
    \includegraphics[width=\columnwidth]{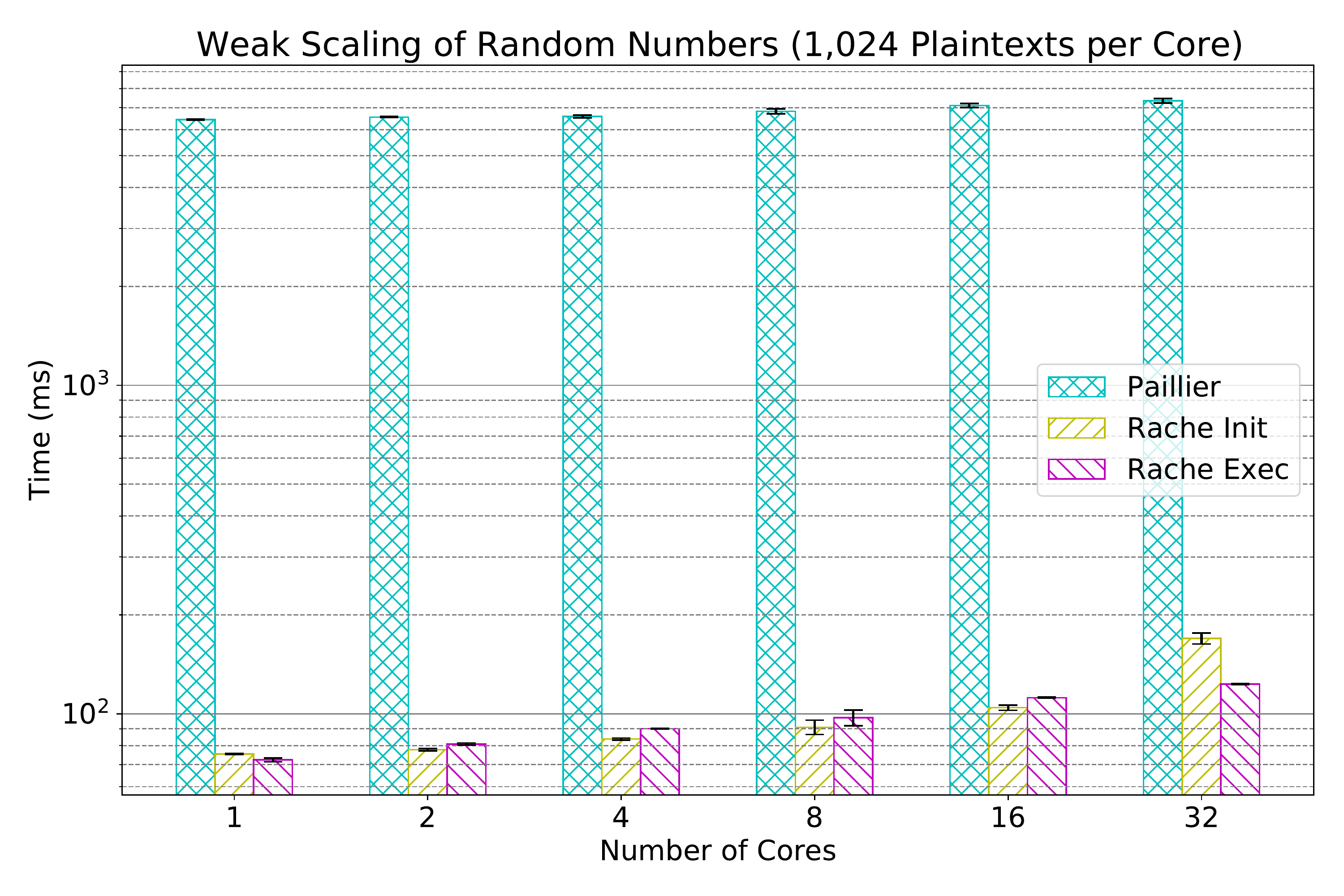}
    \caption{Weak Scaling of Encoding INCHE Random Numbers.}
    \label{fig:weak_scaling}
\end{figure}

Fig.~\ref{fig:weak_scaling} reports the conventional weak-scaling experiment.
We control the workload to be proportional to the number of cores:
1,024 plaintexts for every core.
That is, the workloads range from 1,024 to 32,768 plaintexts of uniformly distributed numbers.
In each workload, the maximal value is roughly the number of plaintexts minus 1 or 2 due to the uniform distribution.
This explains why the Rache overhead (i.e., \textit{Rache Init}) increases proportionally to the number of cores or workloads.

Rache outperforms Paillier by orders of magnitudes at all scales.
However, Rache seems to exhibit a higher slope of encoding time.
We stress that the absolute values of Rache performance are sub-seconds (and the $y$-axis is logarithmic),
therefore the overhead can be best explained by the IPC overhead.
To confirm our conjecture, we conduct the following experiment,
in which we fix the number of cores but increase the workloads.

\begin{figure}[!t]
    \centering
    \includegraphics[width=\columnwidth]{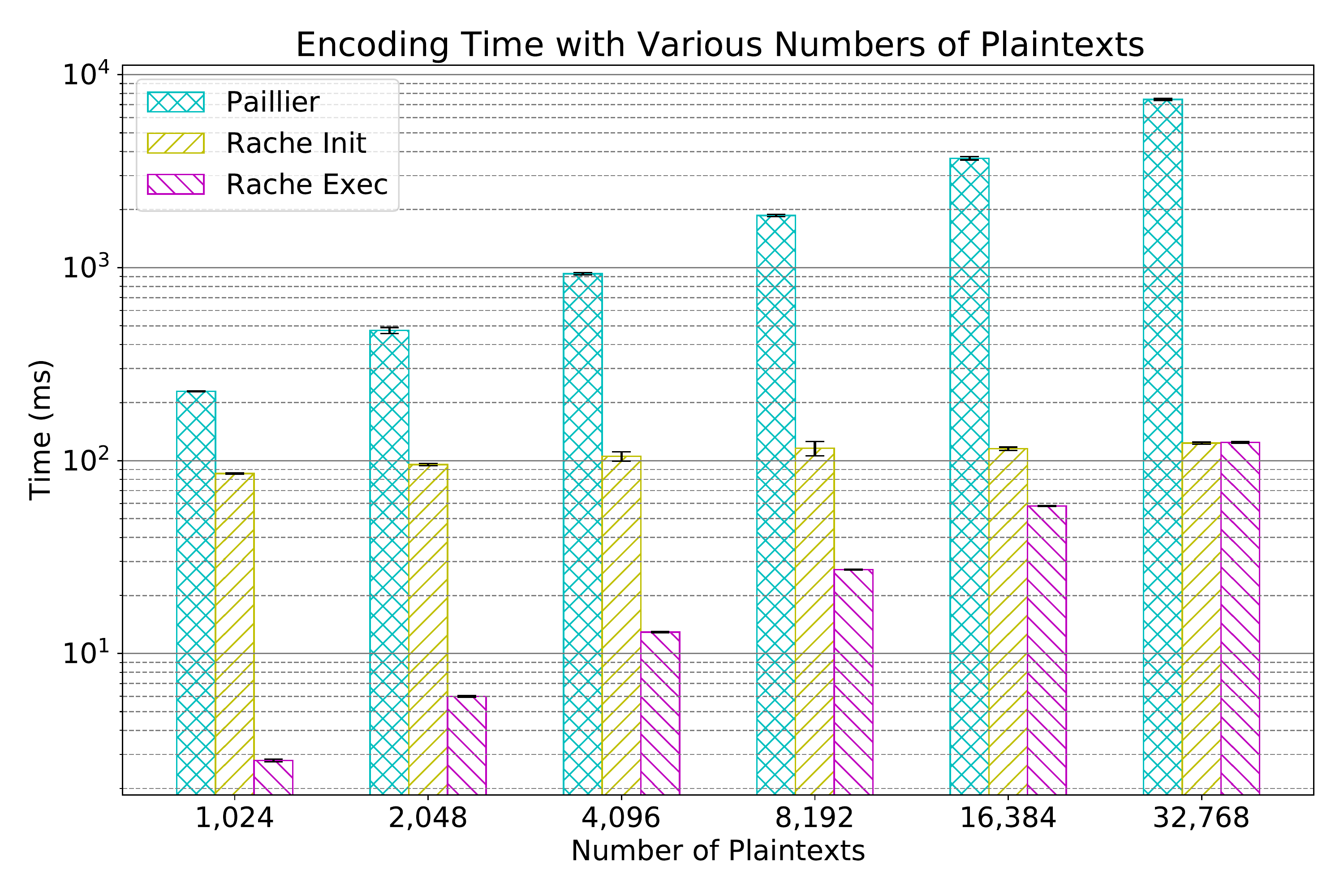}
    \caption{Encoding a variety of workloads with a fix number of cores.}
    \label{fig:fixed_cores}
\end{figure}

Fig.~\ref{fig:fixed_cores} shows the encoding time when we fix the number of cores as 32 but increase the number of plaintexts from 1,024 to 32,768.
We observe that when the IPC overhead is fixed (for 32 cores),
the encoding time is proportionally increased regarding the workload size.
Notably, the Rache initialization overhead is much less noticeable than the previous weak-scaling experiment because the IPC overhead is gone and the only thing remaining is the larger number of radixes when working on more plaintexts. 

\subsection{Applications}
\label{sec:eval_app}

\subsubsection{U.S. Covid-19 statistic}

\begin{figure}[!t]
    \centering
    \includegraphics[width=\columnwidth]{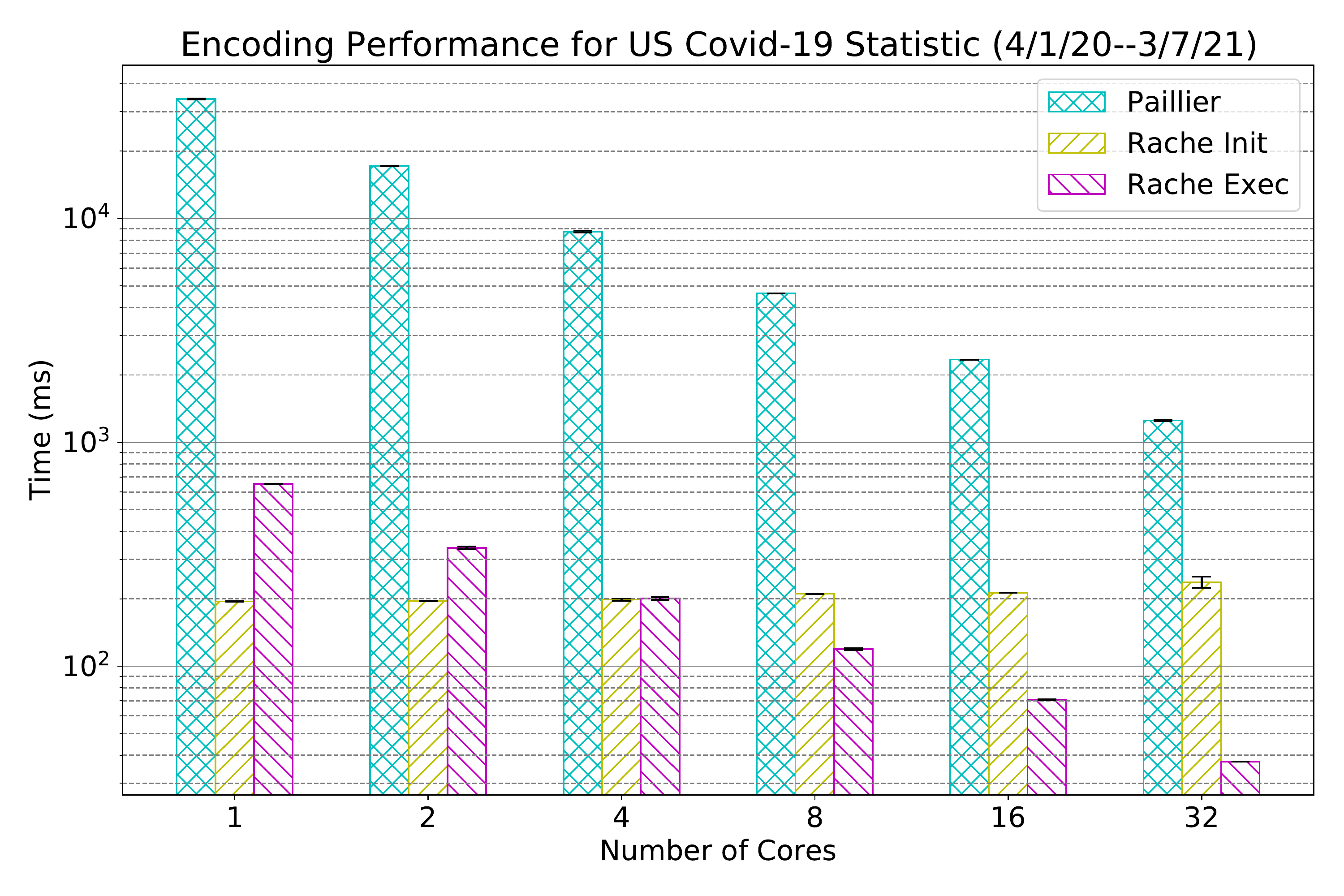}
    \caption{Encoding the U.S. Covid-19 statistic.}
    \label{fig:covid}
\end{figure}

Fig.~\ref{fig:covid} reports the encoding performance of the U.S. Covid-19 statistic published at~\cite{covid19data}.
The data set exhibits a large variety of numbers,
from tens (e.g., number of affected states) to hundreds of millions (e.g., the total number of test results).
This partially impacts the balance between the initialization (i.e., the overhead) and the encoding procedures of Rache:
We observe that with few cores (e.g., 1 and 2) the overhead is smaller than the encoding cost,
while with more cores (e.g., 16, 32) the per-core encoding is very efficient and takes less time than the overhead.
Some of the overhead, i.e., precomputing and caching the large radixes, is unnecessary for those small values,
and yet has to exist due to those extremely large values.
We stress that the overhead is a one-time thing though:
If there were, say, ten years of Covid-19 data,
the overhead would look roughly the same and would be outweighed by the increased cost of encoding the data (cf. Fig.~\ref{fig:fixed_cores}).
As we have seen in the benchmarks, Range outperforms Paillier by almost two orders of magnitude.

\subsubsection{Human genome reference 38}

\begin{figure}[!t]
    \centering
    \includegraphics[width=\columnwidth]{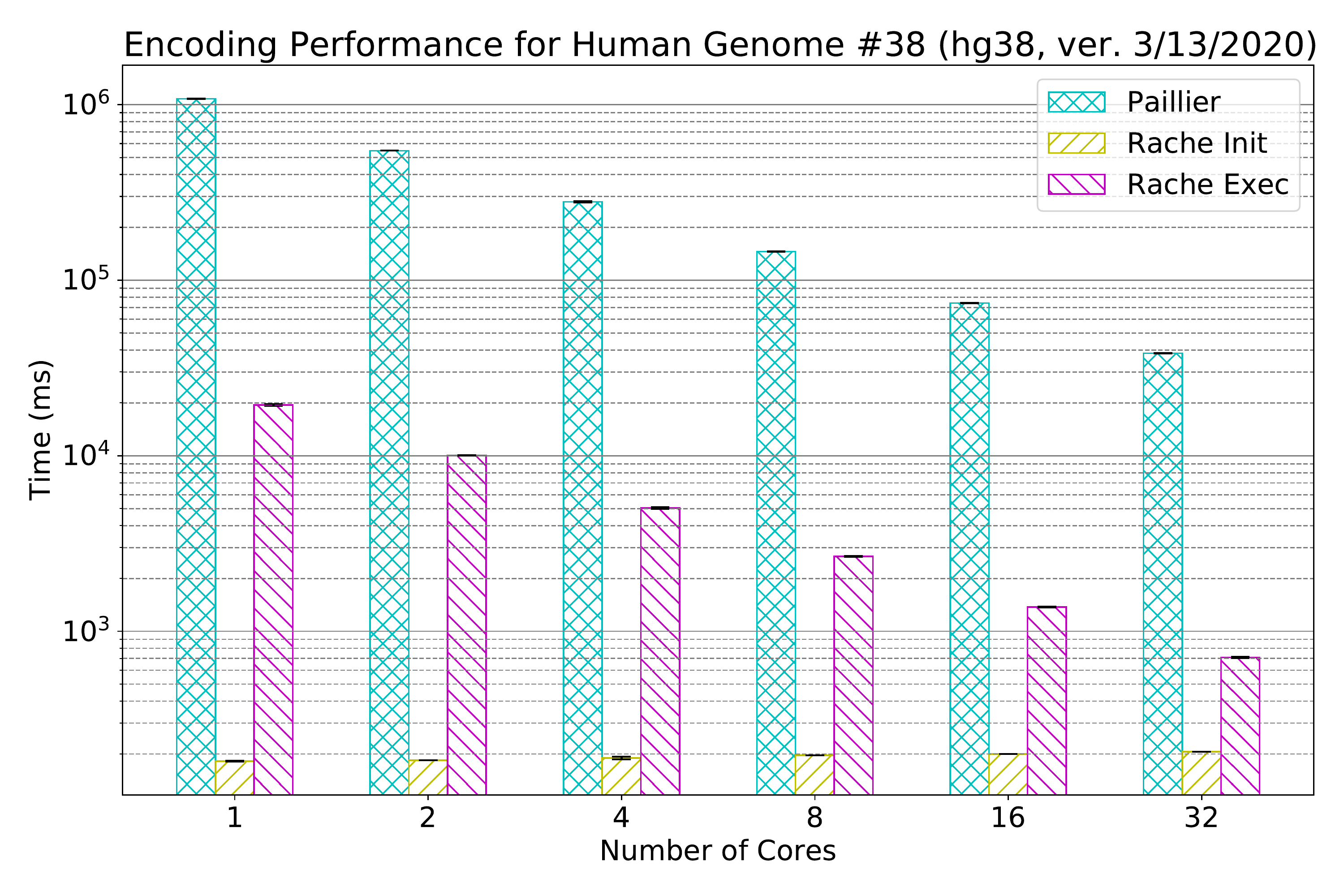}
    \caption{Encoding the human genome reference 38.}
    \label{fig:genome}
\end{figure}

Fig.~\ref{fig:genome} reports the encoding performance of Rache and Paillier on a database of human genome~\cite{hg_data} (\textit{hg38}) that was last updated in March 2020, under the umbrella of the Augustus gene prediction project~\cite{augustus_github}.
As expected, Rache outperforms Paillier at all scales by orders of magnitude.
In sheer contrast to the Covid-19 dataset, 
the initialization overhead of Rache in \textit{hg38} is much less significant:
Even at 32-core, the overhead is less than 30\%.
This is mainly due to a large number of plaintexts (172,120),
whose encoding time greatly outweigh the initialization,
which itself is not trivial either: totally 29 radixes for the largest possible value of 248,937,123.

\subsubsection{Bitcoin trade volume}

\begin{figure}[!t]
    \centering
    \includegraphics[width=\columnwidth]{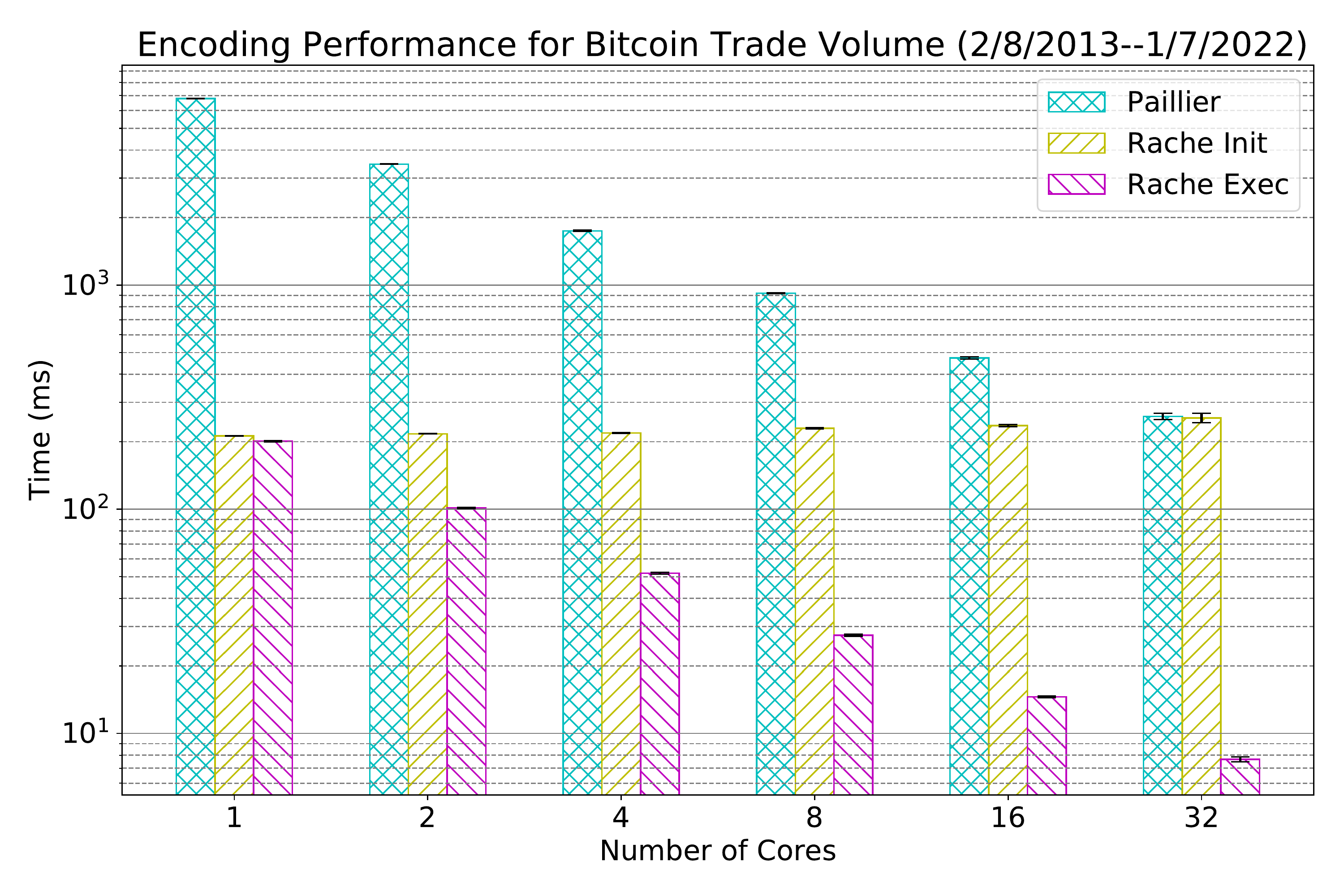}
    \caption{Encoding the Bitcoin trade volume.}
    \label{fig:bitcoin}
\end{figure}

We apply Rache and Paillier to the historical trade volume of Bitcoin exchange since 2013~\cite{bitcoin_trade}.
Fig.~\ref{fig:bitcoin} shows that Rache outperforms Paillier by more than one order of magnitude,
which is consistent with what we have found so far.
The notable thing here is the large overhead incurred by Rache:
on a single core, the overhead is on par with Rache's encoding time;
on 32 cores, the overhead is on par with the Paillier processing time and orders of magnitude larger than Rache's encoding time.
This phenomenon is due to two reasons.
First, the Bitcoin trade volume consists of very large numbers---most are in the order of millions and the largest one is 4,956,849,516 requiring 34 radixes.
Second, the number of plaintexts is relatively small:
there are totally 1,086 plaintexts, each of which records the Bitcoin exchange for the last three days.

\subsection{Memory Footprint}
\label{sec:eval_memory}

\begin{figure}[!t]
    \centering
    \includegraphics[width=\columnwidth]{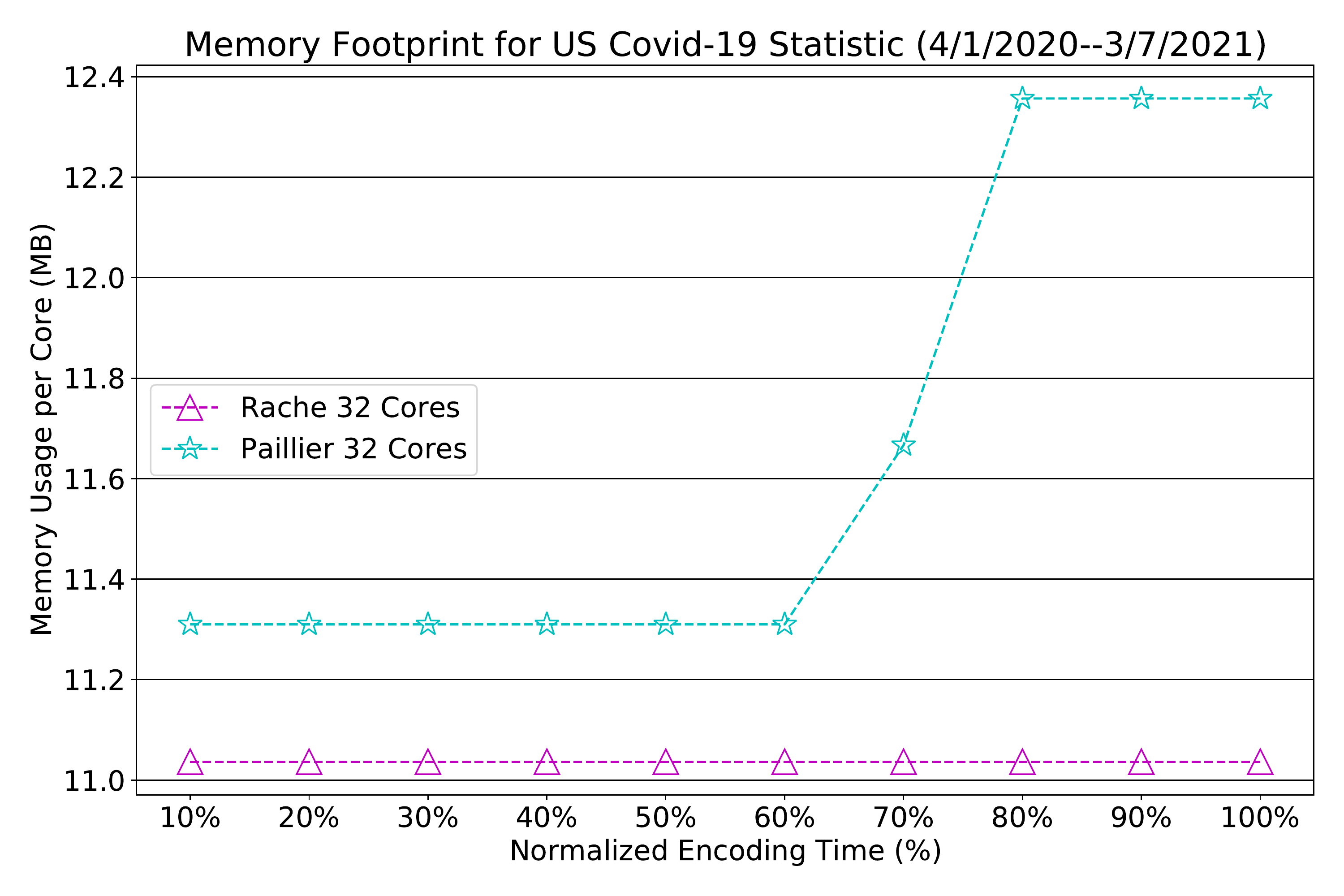}
    \caption{Memory footprint of encoding the U.S. Covid-19 statistic.}
    \label{fig:ram_covid}
\end{figure}

Fig.~\ref{fig:ram_covid} reports the memory footprint of Rache and Paillier when encoding the U.S. Covid-19 statistic.
We only plot the results for the encoding with 32 cores because other scales incur almost the same memory consumption.
It should be noted that the $x$-axis is the normalized timeline:
we probe the memory consumption at every tenth point of the encoding process.
Rache's memory consumption is constant during the encoding process because the main allocation of memory is carried out at the beginning of the process, i.e., initialization of the radixes and the subsequent computations occur directly over the ciphertexts without requiring new memory.
In contrast, Paillier's memory footprint is somewhat sensitive to the plaintexts because they are encrypted on-the-fly and some arbitrarily large plaintexts can cause abrupt allocation of new memory space.
At the end of the encoding procedure, we observe that Rache consumes about 12\% less memory than Paillier, i.e., 11.04 MB vs. 12.36 MB.

\section{Conclusion and Future Work}

This paper presents Rache, a caching optimization for accelerating the performance of homomorphic encryption.
The key insights of Rache are caching some homomorphic ciphertexts before encrypting the large volume of plaintexts and constructing the ciphertexts with only homomorphic addition.
The extensive evaluation shows that Rache exhibits almost linear scalability and outperforms Paillier by orders of magnitude.
    
Our future work will focus on integrating Rache into a blockchain framework called BAASH~\cite{aalmamun_sc21} such that sensitive scientific results can be shared and reproduced in a verifiable manner.
Another follow-up along this line of research is to distribute Rache encoding into a finer granularity,
e.g., topology-aware parallelism~\cite{otawose_ipdps22}.

\bibliographystyle{acm}
\bibliography{ref_new}

\end{document}